\documentclass[sigconf]{acmart}

\AtBeginDocument{%
  \providecommand\BibTeX{{%
    \normalfont B\kern-0.5em{\scshape i\kern-0.25em b}\kern-0.8em\TeX}}}

\setcopyright{acmcopyright}
\copyrightyear{2018}
\acmYear{2018}
\acmDOI{10.1145/1122445.1122456}

\acmConference[Woodstock '18]{Woodstock '18: ACM Symposium on Neural
  Gaze Detection}{June 03--05, 2018}{Woodstock, NY}
\acmBooktitle{Woodstock '18: ACM Symposium on Neural Gaze Detection,
  June 03--05, 2018, Woodstock, NY}
\acmPrice{15.00}
\acmISBN{978-1-4503-XXXX-X/18/06}

\usepackage{graphicx}
\usepackage{subfig}
\begin{document}

\title{Topic, Sentiment and  Impact Analysis: COVID19 Information Seeking on Social Media}

\author{Md Abul Bashar, Richi Nayak, Thirunavukarasu Balasubramaniam}
\affiliation{%
  \institution{Queensland University of Technology}
  \city{Brisbane, Queensland, Australia}}
\email{{m1.bashar, r.nayak, t.balasubramaniam}@qut.edu.au}
\renewcommand{\shortauthors}{Bashar, Richi and Thiru, et al.}

\begin{abstract}
When people notice something unusual, they discuss it on social media. They leave traces of their emotions via text expressions. A systematic collection, analysis, and interpretation of social media data across time and space can give insights on local outbreaks, mental health, and social issues. Such timely insights can help in developing strategies and resources with an appropriate and efficient response. This study analysed a large Spatio-temporal tweet dataset of the Australian sphere related to COVID19. The methodology included a volume analysis, dynamic topic modelling, sentiment detection, and semantic brand score to obtain an insight on the COVID19 pandemic outbreak and public discussion in different states and cities of Australia over time. The obtained insights are compared with independently observed phenomena such as government reported instances.

\end{abstract}



\maketitle

\section{Introduction}
An outbreak of infectious diseases such as COVID19 has a devastating impact on society with severe socio-economic consequences. The COVID19 pandemic has already caused the largest global recession in history;
global stock markets have crashed,
travel and trade industries are losing billions, schools are closed, and health care systems are overwhelmed. Mental health and social issues creep up as people fear of catching the disease or losing loved ones, as they lose jobs, or as they are required to stay in isolation.

An insight into an outbreak is essential for controlling infectious diseases and identifying subsequent mental and social issues \cite{Al-garadi2016UsingReview}. This will help in reducing costs to the economy over the long term and bringing harmony to the society.
Especially, early detection helps in placing strategies and resources for an appropriate and efficient response. 
On social media, people discuss things that they observe in community \cite{Al-garadi2016UsingReview} 
They leave traces of their emotions via text expressions \cite{Gkotsis2017CharacterisationLearning}. 
A systematic collection, analysis, and interpretation of social media data can give insight into an outbreak. 
Twitter is one of the most popular micro-blogging social media websites where users express their thoughts and opinions on real-world events\cite{Dahal2019TopicTweets}. 
Social scientists have used tweet datasets for various purposes such as investigating public opinion of Hurricane Irene \cite{Mandel2012AIrene} and election result prediction \cite{TumasjanAndranikandSprengerTimmOandSandnerPhilippGandWelpe2004PredictingSentiment}. 

Recently, the spatio-temporal texts collected from Sina-Weibo (Twitter alike microblogging system in China) was analysed to understand public opinions on COVID19 related topics \cite{Han2020UsingChina}. The static topic modelling technique (LDA) and Random Forest classifier were used to group tweets into topics for analysis. 
Studies have also been published to analyse climate change-related tweets and understand what are the topics of discussion, how the tweet volume and sentiment changed over time \cite{Abdar2020EnergyTweets,Ballestar2020TheApproach,Dahal2019TopicTweets}.
Authors in \cite{Lansley2016TheLondon} applied topic modelling on a corpus of geotagged tweets collected from the London sphere. Topic modelling has also been used to estimate the similarity between users in a location-based social network \cite{Lee2016ANetwork} and to estimate the relatedness of businesses based on business descriptions \cite{Shi2015TowardIntelligence}. 

In this paper, we analyse a large Spatio-temporal tweet dataset of the Australian sphere Twitter\footnote{Location of author or tweet or a location mentioned in the tweet is Australia or any of its cities} containing certain keywords relating to COVID19. The methodology included a volume analysis, Dynamic Topic Modelling \cite{Blei2006DynamicModels}, Sentiment Detection \cite{Medhat2014SentimentSurvey},
and Semantic Brand Score (SBS) \cite{FronzettiColladon2018TheScore} 
to obtain an insight into COVID19 outbreak in different states and cities of Australia over the time. 

The obtained insights are compared with independently observed phenomena such as government reported instances and news on newspapers. To our knowledge, ours is a first in-depth study of understanding Australian people's perception of this ongoing COVID19 pandemic using a large Twitter data collection. More specifically, this study makes the following main contributions.
(a) Investigates how closely the insights into local outbreak match independently observed phenomena in space and time.
(b) Understands what topics related to COVID19 have been discussed in communities and how they change over time. 
(c) Understands the COVID19 related sentiments in communities over time.
(d) Investigates the impact of COVID19 related concepts or words on social media discussion.
(e) Proposes a simple but effective CNN architecture for sentiment analysis.

\section{Experimental Methodology}
\label{sec:exp_methodology}
\subsection{Methodology}
The aim of this study is to use social media analysis to uncover what is happening in communities and to give insight into (a) how the virus and lockdown is affecting community emotions, (2) understand the main topics or themes emerging and evolving in the conversation and (3) impact of different COVID19 related concepts.

In this study, we conduct spatio-temporal analysis of volume, sentiment, topic, and impact to a large volume of COVID19 related tweets from the Australian sphere. 
First, we collect a dataset of tweets from the Australian sphere containing geospatial and temporal values. The dataset is then preprocessed and prepared for volume analysis, sentiment analysis, dynamic topic modelling, and impact analysis. 
The volume analysis aims to identify basic geospatial and temporal facts from the dataset which will facilitate subsequent analysis such as sentiment and topic into context. Dynamic topic modelling extracts topics present in the dataset and shows how those topics evolve over time. Sentiment analysis determines the sentiment of every tweet to show how community sentiments change over time. Impact analysis generates networks of concepts or words from the text collection and uses those networks to measure how differently the concepts or words impact a discussion. The analytical findings are then discussed and evaluated along with the comparison with independent observations.

\begin{figure}
    \centering
    \includegraphics[width=0.3\textwidth]{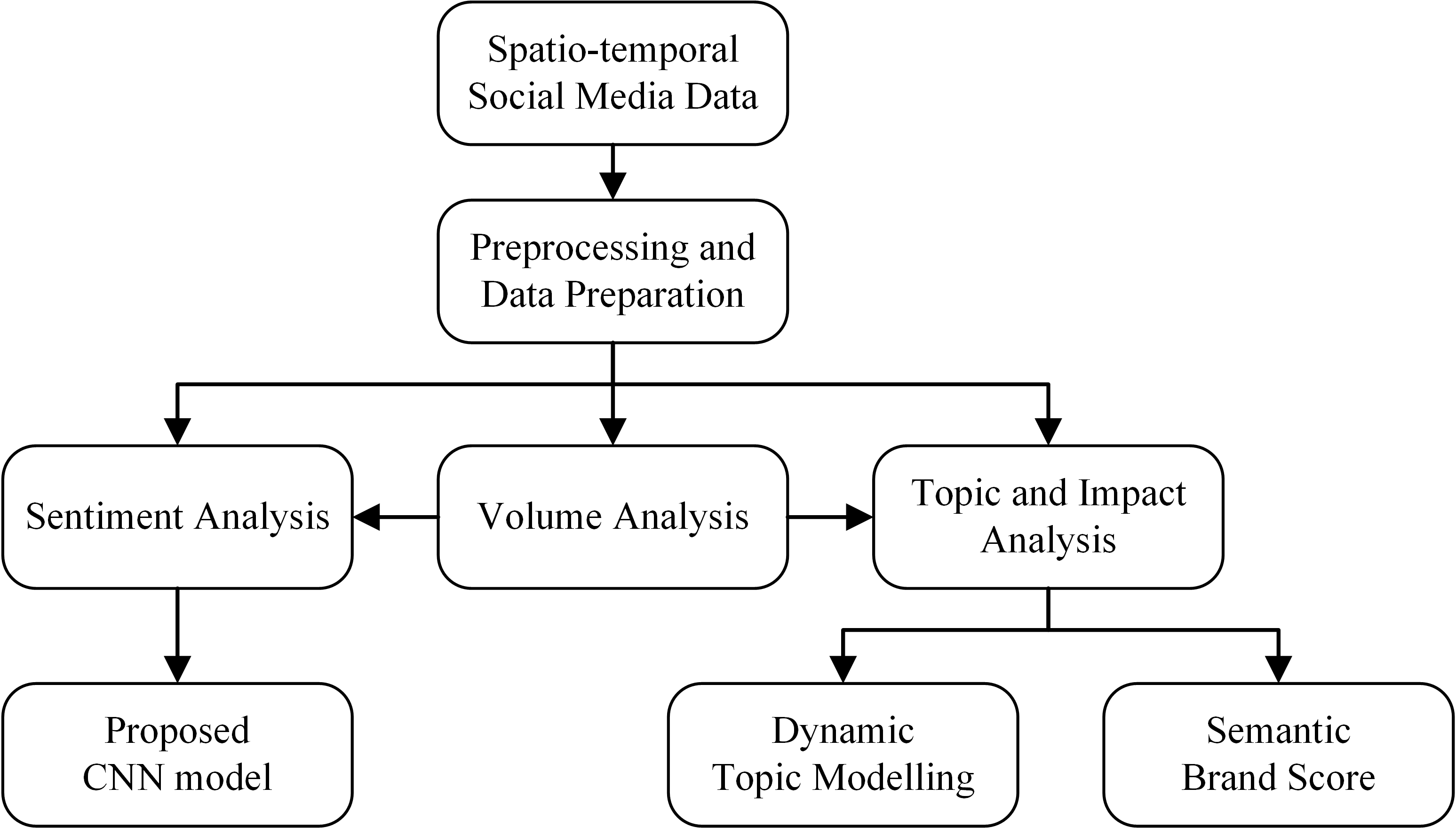}
    \caption{The Experimental Workflow}
    \label{fig:workflow}
\end{figure}

\vspace{-2mm}
\subsection{Data Collection and Preparation}
As we practice social distancing, our embrace of social media gets only higher. The major social media platforms have emerged as critical information purveyors during the expanding pandemic. Twitter’s number of active users in the first three months of 2020 increased by 23\% compared to the end of 2019, which is about 12 million more users. 

We collected Twitter conversation in the Australian Sphere on Coronavirus since November 27th when the first break out occurred in China. The data collection is done via the QUT facility of Digital Observatory\footnote{https://bit.ly/2Z6RUvU} using the Twitter Stream Application Programming Interface (API). The dataset consists of 2.9 million tweets from 27 November 2019 to 7 April 2020. Every tweet in the dataset contains or uses as a hashtag at least one of the following keywords: coronavirus, covid19, covid-19, covid\_19, coronovirusoutbreak, covid2019, covid, and coronaoutbreak. 

The body of each tweet, i.e. tweet message, is used for analysing sentiment, topics and impact. Location and time information of each tweet gives it spatio-temporal dimensions. The location information for each tweet comes from either of three sources based on their availability: (a) tweet location, i.e. the user was in when the tweet was posted; (b) user location, i.e. residence of the user; or (c) a location mentioned in the tweet message. The locations are mapped to capital cities, states, or the country Australia depending on how granular level location information could be extracted. The time information of each tweet comes from the time and day the tweet was posted. Table \ref{tab:tweet_samples} shows a few examples.

For preprocessing, we removed stopwords, punctuations, invalid characters. We dropped any non English tweets. We fixed repeating characters, converted text to lowercase, replaced an occurrence of link or URL with a token namely xurl, and stemmed the text. 
We used Named-entity Recognition (NER) technique from spacy\footnote{https://spacy.io/usage/linguistic-features} to extract locations.

For sentiment analysis, we collected 
the sentiment140 dataset\footnote{https://www.kaggle.com/kazanova/sentiment140} from kaggle. This dataset contains 1.6 million annotated tweets. The tweets are annotated for classes of sentiments: positive and negative. We train a classifier model using these tweets to detect sentiment in the collected dataset of 2.9 million tweets. We use the same preprocessing as above to prepare this dataset.

\begin{table*}
  \centering
  \scriptsize
  \caption{Example of tweets in the dataset. Location is extracted, @someone and @something is used to anonymise a person or an organisation mentioned in the tweet, a token URL is used to replace any occurrence of hyperlink or URL.}
    \begin{tabular}{p{1.5cm}p{10cm}p{2.1cm}}
    \toprule
    Location & Tweet Text & Time \\
    \midrule
    Australia & RT @someone: Coronavirus patient sealed in a PLASTIC TUBE to avoid contamination URLl @something & 22/01/2020 17:47 \\
    Melbourne & RT @someone: Me seeing the doomsday clock going to a 100 seconds, Australia on fire and the coronavirus all trending on the same day \textbackslash{}n \#AustraliaOnFire\textbackslash{}n \#CoronavirusOutbreak\textbackslash{}n \#DoomsdayClock\textbackslash{}n \#Wuhan\textbackslash{}n \#coronovirus URL & 24/01/2020 0:50 \\
    Tasmania & RT @someone: BREAKING: virologist who helped identify SARS says a bigger \#CoronavirusOutbreak is certain, conservatively estimating it could be 10x bigger than SARS because SARS was transmitted by only a few super spreaders in a more defined part of \#China.\textbackslash{}n URL & 24/01/2020 13:26 \\
    \bottomrule
    \end{tabular}%
  \label{tab:tweet_samples}%
\end{table*}%

\vspace{-2mm}
\subsection{Volume analysis}
Analysing the volume of tweets posted from a particular area at a particular time is an important step of data exploration that can provide interesting insights into observations \cite{Dahal2019TopicTweets}. We analysis the number of tweets posted in each state and capital in Australia over time. 

\vspace{-2mm}
\subsection{Sentiment analysis}
Sentiment analysis is used to identify the emotional state or opinion polarity in the samples. 
We identify the sentiment of each tweet using our proposed CNN based sentiment classifier. We then aggregate tweets by location and time to obtain spatio-temporal distribution of sentiments. The following section gives a summary of the proposed CNN architecture for sentiment classification. 


\subsubsection{Model Architecture}
\begin{figure*}[ht]
    \centering
    \includegraphics[width=0.5\textwidth]{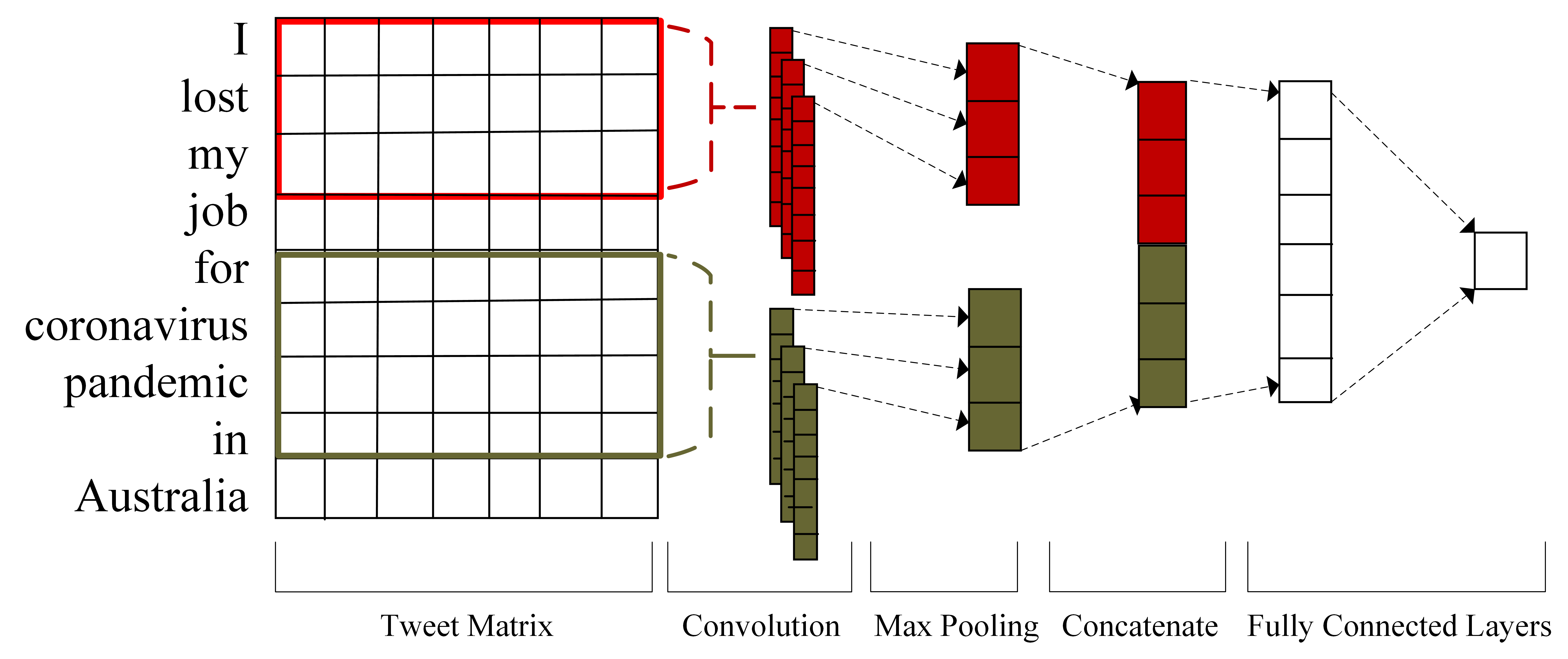}
    \caption{CNN Model for Sentiment Identification}
    \label{fig:ModelArchitecture}
\end{figure*}

Figure \ref{fig:ModelArchitecture} illustrates the architecture of CNN model used to identify sentiments of COVID19 related tweets. 
The model uses word embedding \cite{Bashar2019QutNocturnalHASOC19:Language,mikolov2013efficient} to represent each word $w$ in an $n$-dimensional word vector $\mathbf{w} \in \mathbb{R}^n$ where the dimension $n$ is empirically set to 200. We represent a tweet $t$ with $m$ words as a matrix $\mathbf{t} \in \mathbb{R}^{m \times n}$ utilising word embedding. We apply the convolution operation to the tweet matrix with one stride. 

Each convolution operation applies a filter $\mathbf{f}_i \in \mathbb{R}^{h \times n}$ of size $h$. The convolution is a function $\mathbf{c}(\mathbf{f}_i, \mathbf{t}) = r(\mathbf{f}_i \cdot \mathbf{t}_{k:k+h-1})$, where $\mathbf{t}_{k:k+h-1}$ is the $k$th vertical slice of the tweet matrix from position $k$ to $k+h-1$, $\mathbf{f}_i$ is the given filter and $r$ is a Rectified Linear Unit (ReLU) function.
The function $\mathbf{c}(\mathbf{f}_i, \mathbf{t})$ produces a feature $c_k$ similar to $n$Grams for each slice $k$, resulting in $m-h+1$ features. We apply the max-pooling operation 
over these features and take the maximum value, i.e. $ \hat{c}_i = max(\mathbf{c}(\mathbf{f}_i, \mathbf{t}))$. Max-pooling captures the most important feature for each filter. Empirically, based on the accuracy improvement in ten-fold cross validation, 256 filters are used for $h \in \{3,4\}$ and 512 filters for $h \in \{5\}$. As there are a total of 1024 filters (256+256+512) in the proposed model, the 1024 most important features are learned from the convolution layer. 

We then pass these features to a fully connected hidden layer with 256 perceptrons that use the ReLU activation function. This fully connected hidden layer learns the complex non-linear interactions between the features from the convolution layer and generates 256 higher-level new features. Finally, we pass these 256 higher-level features to the output layer with a single perceptron that uses the sigmoid activation function. The perceptron in the output layer generates the probability of the tweet in our data collection being positive or negative sentiment. 

In this architecture (Figure \ref{fig:ModelArchitecture}), a proportion of units are randomly dropped-out from each layer except the output. This is done to prevent the co-adaptation of units in a layer and to reduce overfitting. 
We set 50\% units dropout in the input layer, the filters of size 3, and the fully connected hidden layer based on best empirical results. Only 20\% units are set to dropout from the filters of sizes 4 and 5. Python code for this model is available online at https://github.com/mdabashar/sentiment\_analysis.

\vspace{-4mm}
\subsection{Topic Analysis}
A variety of subjects or topics are usually discussed in the tweets over time. Knowing those topics and how they evolve is important to understand the dynamics of discussion related to coronavirus. 
Because of the large size of the tweet dataset, it is very difficult, if not impossible, to read all of the tweets for finding out their topics. We use an unsupervised machine learning technique known as topic modelling \cite{Blei2006DynamicModels} to discover subjects topics of discussion in tweets and how those topics evolve over time. Topic models are the most popular statistical methods that analyse the words in a document collection to discover the themes that run through the documents, how those themes are connected, and how they change over time \cite{Blei2006DynamicModels,blei2003latent}. 

In general, a topic modelling technique (e.g. Latent Dirichlet Allocation \cite{blei2003latent}) uses word co-occurrences within documents for finding topics in a document collection. Words occurring in the same document are more likely coming from the same topics \cite{blei2003latent,Dahal2019TopicTweets}; and documents that contain the same words are more likely consist of the same topics \cite{blei2003latent,Dahal2019TopicTweets}. We use each tweet as a document to discover topics in our tweet collection. 


Static topic modelling \cite{blei2003latent}  explicitly treats words exchangeably and implicitly treats documents exchangeably \cite{Blei2006DynamicModels}. 
However, the assumption of exchangeable documents is inappropriate for many collections such as tweets, news articles, and scholarly articles as they are evolving content. For example, tweets published in different time periods may be related to a specific topic namely \emph{coronavirus cure} but the coronavirus cure can be much different in later stages of time than the early stage. The themes in a tweet collection evolve, and it is of interest to explicitly model the dynamics of the underlying topics. 

Dynamic topic modelling \cite{Blei2006DynamicModels} extends static topic modelling \cite{blei2003latent} to incorporate topic evolution. Dynamic topic modelling can capture the evolution of topics in a sequentially organised collection of tweets or documents. In this setting, tweets are grouped by weeks, and each week’s tweets arise from a set of topics that have evolved from the last week’s topics. We use dynamic topic modelling to observe topics of discussion and how they change over time.

Choosing a reasonable number of topics is important because too few topics could lead to merging distinct topics whereas too many topics could result in fragmented topics that otherwise could make a cohesive topic together. Therefore, we manually evaluated topic models with topics ranging from 5 to 50, to determine the optimal number of topics. Other hyperparameters are set to the default value in gensim (the Python software library used for topic modelling).

To ensure that the topics discovered by dynamic topic modelling are meaningful and not dominated by the same top words, we removed keywords and hashtags (e.g. covid19, coronavirus, etc.) that we used for collecting our tweets. We also removed rare words (i.e. words with a very low frequency) to reduce noise in the topics.

\vspace{-3mm}
\subsection{Impact Analysis}
The Semantic Brand Score (SBS) estimates the impact or importance of concepts, brands, or entities in a text collection \cite{FronzettiColladon2018TheScore}. We use SBS to understand the impact of different COVID19 related concepts or entities on social media discussion. 

SBS is based on graph theory that combines methods of Social Network and Semantic Analysis using the word co-occurrence network. It is estimated as the standardised sum of three components: prevalence, diversity, and connectivity. 

Prevalence $PREV(c)$ calculates the number of times a word or concept $c$ is mentioned in a text collection. Prevalence is associated with the idea of brand awareness assuming that when a concept is frequently mentioned increases its recognition and recall for those who read it. 

Diversity $DIV(c)$ of a word or concept $c$ estimates the heterogeneity of concepts surrounding the concept $c$. It is the degree of centrality in the co-occurrence network. The degree of centrality is estimated by counting the number of edges directly connected to the concept node $c$. 

Connectivity $CON(c)$ of a word or concept $c$ estimates the connectivity of the concept $c$ with respect to a general discourse. It represents the ability of the concept node $c$ to act as a bridge between other nodes in the network. Connectivity is widely used in social network analysis as a measure of influence or control of information that goes beyond direct links. It is estimated as 
\[
CON(c) = \sum_{j \neq k} \frac{d_{jk}(c)}{d_{jk}}
\]
where $d_{jk}$ is the number of the shortest paths linking any two nodes $j$ and $k$, and $d_{jk}(c)$ is the number of those shortest paths that contain the given concept node $c$. 

The Semantic Brand Score is estimated as 
\[
SBS(c) = \frac{PREV(c)-\overline{PREV}}{std(PREV)} + \frac{DIV(c)-\overline{DIV}}{std(DIV)} + \frac{CON(c)-\overline{CON}}{std(CON)}
\]
where $\overline{.}$ represents the mean value and $std$ represents the standard deviation. 

\vspace{-2mm}
\section{Experimental results}
\label{sec:results_duscussion}
The temporal dimension (27 November 2019 to 7 April 2020) of the tweet collection is discredited by weeks (roughly 17 weeks) or days as appropriate to the nature of the analysis. The geospatial dimension is discredited by Australian Sates and capital cities. The tweet user location that does not list city but lists the country as Australia is categorised as Australia (au). The locations of a small portion of tweets could not be extracted or mapped to our selected categories that we categorise as others (oth). 

\vspace{-2mm}
\subsection{Volume Analysis}
\begin{figure}%
    \centering
    \includegraphics[width=0.25\textwidth]{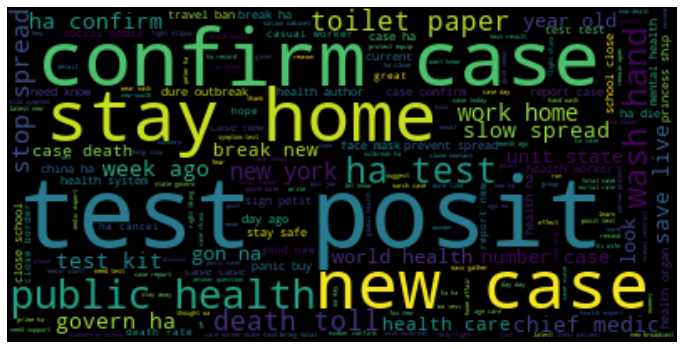}%
    \caption{A Word Cloud Generated from the entire Tweet Collection}%
    \label{fig:word_cloud_full}%
\end{figure}

Figure \ref{fig:word_cloud_full} shows a word cloud generated from the entire tweet collection. 
It gives a quick look into the subjects Australian people are discussing during the COVID19 pandemic. Subjects such as `stay home', `work from home', `toilet paper crisis' `slow the spread' etc. are commonly discussed.

\begin{figure}%
    \centering
    \includegraphics[width=0.3\textwidth]{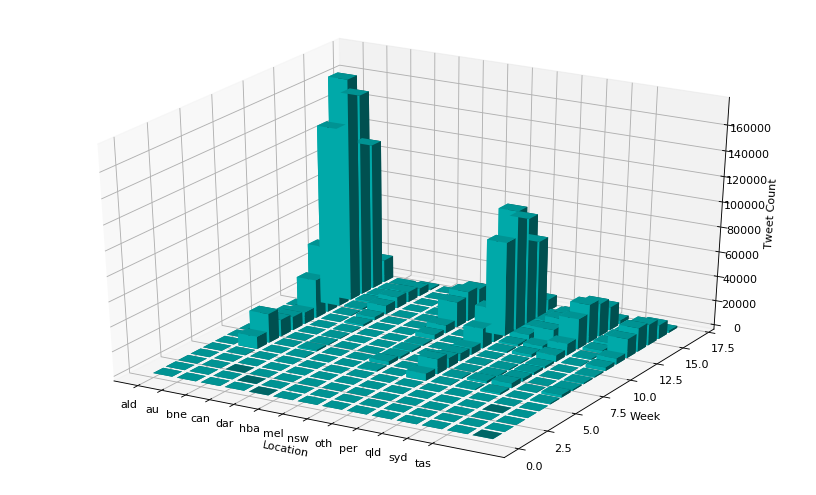}%
    \caption{Geospatial and Temporal Distribution of Tweet Count}
    \label{fig:spatio_temporal_tweet_count}%
\end{figure}

\begin{figure}[htb!]
    \centering
    \subfloat[Tweet Count Distributed over States, Territories and Capital Cities]{{\includegraphics[width=4cm]{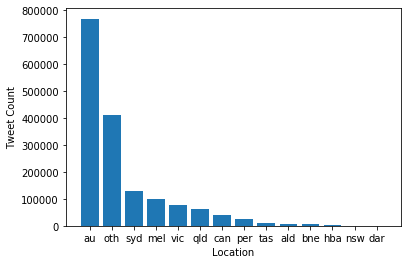}}\label{fig:Compare_count_case_space_a}}
    \qquad
    \subfloat[COVID19 Cases Distribute over States and Territories \cite{HealthCoronavirusNumbers}]{{\includegraphics[width=4cm]{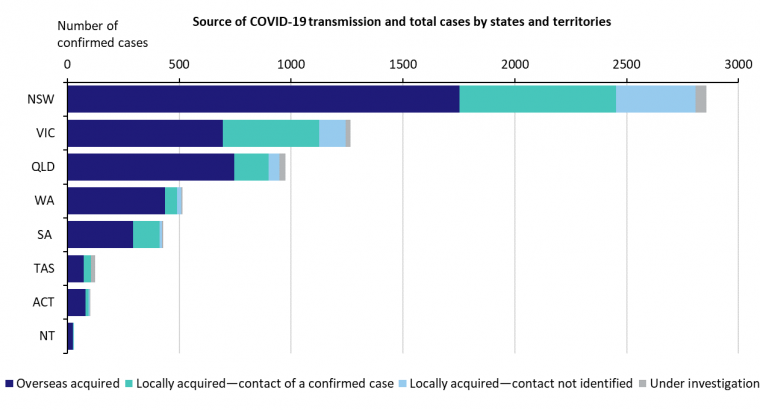} }\label{fig:Compare_count_case_space_b}}%
    \caption{Correlation between \emph{Tweet Counts} and \emph{COVID19 Cases} Distributed over States, Territories and Capital Cities}%
    \label{fig:Compare_count_case_space}%
\end{figure}

\begin{figure}[htb!]
    \centering
    \subfloat[Tweet Count Distributed over Time]{{\includegraphics[width=4cm]{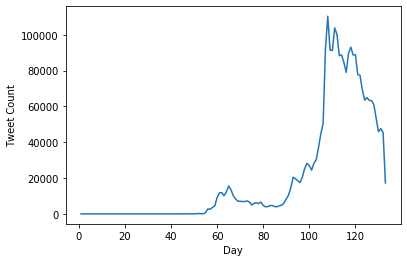}}\label{fig:Compare_count_case_time_a}}\hfill%
    \qquad
    \subfloat[COVID19 Cases Distribute over Time \cite{HealthCoronavirusNumbers}]{{\includegraphics[width=4cm]{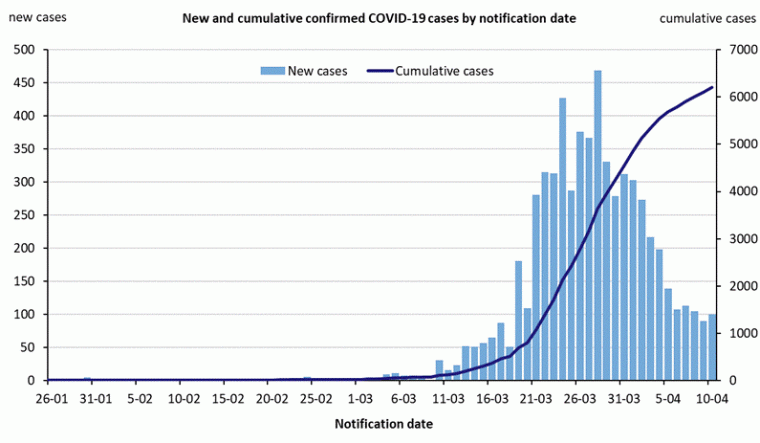}}\label{fig:Compare_count_case_time_b}}%
    \caption{Correlation between \emph{Tweet Counts} and \emph{COVID19 Cases} Distributed over Time}%
    \label{fig:Compare_count_case_time}%
\end{figure}

Figure \ref{fig:spatio_temporal_tweet_count} shows the geospatial and temporal distribution of tweet count. The significant changes in tweet counts over the locations and weeks can be noted throughout the time period. For a closer examination, we separate geospatial and temporal dimensions in Figures \ref{fig:Compare_count_case_space} and \ref{fig:Compare_count_case_time} respectively.

Figure \ref{fig:Compare_count_case_space_a} shows the number of tweet counts in states, territories and capital cities in Australia. Figure \ref{fig:Compare_count_case_space_b} shows the actual number of COVID19 positive cases in states and territories in Australia. A strong correlation can be noted between tweet counts and COVID19 cases. The more the number of COVID19 cases in a location, the more the number of tweets there. For example, the highest number of COVID19 related tweets were observed in Sydney (syd) (i.e. the capital city of New South Wells (NSW)), where the highest number of COVID19 cases occurred in NSW. The second and the third-highest number of COVID19 related tweets were observed in Melbourne (mel) (i.e. the capital city of Victoria (VIC)) and VIC respectively, where the second-highest number of COVID19 cases occurred in VIC. The same is true for Queensland (QLD). Other cities follow a similar pattern with minor order variations.

Fig \ref{fig:Compare_count_case_time} shows the correlation between tweet counts and COVID19 cases distributed over time. A comparison between Figures \ref{fig:Compare_count_case_time_a} and \ref{fig:Compare_count_case_time_b} shows that the total number of COVID19 related tweets over time is strongly correlated with the number of new COVID19 positive cases by the notification date.

Figure \ref{fig:Compare_count_case_time_a} shows that when COVID19 hit China on 27 November 2019, there were not many discussions held in Australian space. A noticeable number of coronavirus related tweets started to be posted after 60 days or around eight weeks, i.e. end of January. Next one week the number increased and then started to fall. The main burst of tweets started after another 30 days or 4 weeks, i.e. the end of February. This might be because this time several people in Australia from overseas were identified COVID19 positive. The number exponentially increased for the next 20 days and reached its peak by the third quarter of March. This exponential increase might have occurred because during this time many Australian were identified COVID19 positive and some of them were reported dead. Then it started to fall gradually. This might be because during this time government initiatives and strict social distancing worked and the COVID19 infection death rate started to decrease.

\vspace{-5mm}
\subsection{Sentiment Analysis}
The experimental performance comparison of the proposed CNN model and the pretrained language model ULMFiT \cite{howard2018universal} is given in Table \ref{tab:compare_cnn_ULMFiT}. ULMFiT is pretrained with  Wikitext-103 that contains 28,595 preprocessed Wikipedia articles and 103 million words \cite{merity2016pointer}. The model is then finetuned with 1.6 million tweets from the sentiment1040 dataset without labels. Finally, we add an extra classifier layer at the end of the model and train the model with the 1.6 million tweets from the sentiment1040 dataset with labels. We use the same architecture, hyper parameters and training strategy for ULMFiT as described in  \cite{howard2018universal}. We use 80\%, 10\%, and 10\% split of data for training, validation, and testing respectively. 

Table \ref{tab:compare_cnn_ULMFiT} shows that the proposed CNN achieves similar performance as of ULMFiT. The performance of the CNN-based model shows that a carefully designed simple model can achieve a similar performance of a sophisticated model when a reasonably sized training dataset is available. 
The significance of this finding is that sophisticated pretrained language models, such as ULMFiT, are computationally expensive and memory intensive. Effectively using them becomes difficult for monitoring (i.e. classifying) a large tweet collection in a resource-constrained environment commonly available to practitioners. A simple model, such as our posed CNN, that achieves a similar performance can greatly help in this regard.

\vspace{3mm}
\begin{table}[htbp]
  \centering
  \caption{Performance Comparison of ULMFiT and the Proposed CNN Model}
    \begin{tabular}{lcc}
    \toprule
          & CNN   & ULMFiT \\
    \midrule
    True Positive & 63465 & 65144 \\
    True Negative & 63399 & 62300 \\
    False Positive & 16407 & 17738 \\
    False Negative & 16729 & 14808 \\
    Accuracy & 0.793 & 0.797 \\
    Precision & 0.795 & 0.786 \\
    Recall & 0.791 & 0.815 \\
    F$_1$ Score & 0.793 & 0.800 \\
    Cohen Kappa & 0.586 & 0.593 \\
    Area Under Curve & 0.793 & 0.797 \\
    \bottomrule
    \end{tabular}%
  \label{tab:compare_cnn_ULMFiT}%
  \vspace{-4mm}
\end{table}%

\begin{figure}[htb!]%
    \centering
    \includegraphics[width=0.25\textwidth]{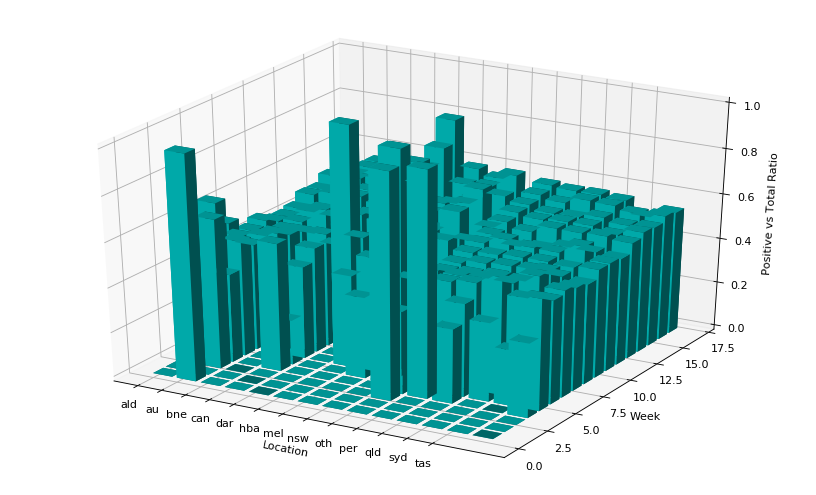}%
    \caption{Geospatial and Temporal Distribution Number of Positive vs Total Tweet Ratio}%
    \label{fig:spatio_temporal_positive_vs_total_ratio}%
\end{figure}

The following results of sentiment analysis are based on the proposed CNN-based model. Figure \ref{fig:spatio_temporal_positive_vs_total_ratio} shows geospatial and temporal distribution of positive sentiment tweet counts vs total tweet counts for an overall observation. As soon as COVID19 hit the world, the positive sentiments dropped sharply (from roughly 85\% to roughly 48\% on average). The percentage stayed there for up to around 12 weeks. Then it gradually changed for three weeks with a very marginal increment. For the final two weeks, the increment was a bit more than the previous three weeks. 

A possible explanation of the trend can be explained as follows. As soon as COVID19 hit the world, the online community got shocked by the news. It took some time for world leaders to come up with plans on how to combat COVID19. During this period (12 weeks) people remained stressed. When the world leaders explained their combat plans and ideas, twitter users talked about those positive initiatives during this time period (three weeks). During the final two weeks, the Australian government announced social safety plans, e.g. economic aids to organisations, businesses, and individuals; it announced more strict rules for social distancing and the COVID19 infection curve started flattening. People started to become a bit comfortable and discussed these positive aspects in their tweets. Consequently, the number of positive tweets has increased. All these patterns show that by monitoring conversational dynamics on social media, we can identify how people are feeling during the pandemic of COVID19, and what initiatives are working or making people comfortable. 

\begin{figure}[htb!]%
    \centering
    \includegraphics[width=0.23\textwidth]{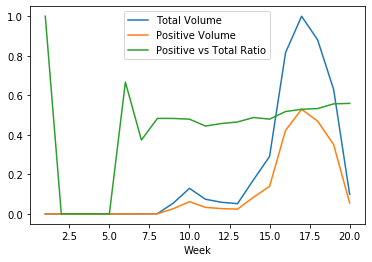}%
    \caption{Temporal Distribution of Positive and Total Volume of Tweets}%
    \label{fig:temporal_sentiment}%
\end{figure}

\begin{figure}[htb!]
    \centering
    \subfloat[Geospatial Distribution of Positive vs Total Number of Tweet Ratio]{{\includegraphics[width=4cm]{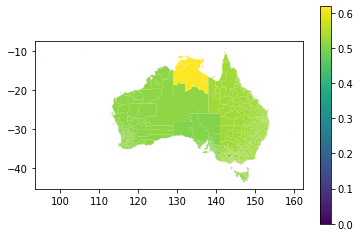}}\label{fig:geospatial_sentiment_a}}\hfill%
    \qquad
    \subfloat[Geospatial Distribution of Positive and Total Number of Tweets]{{\includegraphics[width=4.5cm]{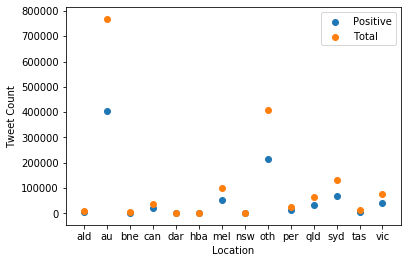}}\label{fig:geospatial_sentiment_b}}%
    \caption{Geospatial Distribution of Sentiment}%
    \label{fig:geospatial_sentiment}%
\end{figure}

For a closer look into the trend, the geospatial and temporal dimensions are decoupled in Figure \ref{fig:geospatial_sentiment} and \ref{fig:temporal_sentiment} respectively. Figure \ref{fig:temporal_sentiment} shows the volume of COVID19 related tweets (total volume), the volume of positive tweets related to COVID19 (positive volume), and their ratio (positive vs total ratio). This figure shows that, roughly at any time, among all the COVID19 related posts, only 50\% of them were positive. We see two significant drops in the ratio of positive sentiments, one is at the beginning when the world was hit by COVID19 and the next one is by week seven or third quarter of January 2020. During this period there were not many discussions of COVID19 in Australia. However, the second drop triggered an increase in the number of COVID19 related posts. In other words, this second drop alerted the community about the upcoming danger of COVID19. We can assume that the small number of tweets related to COVID19 might come from the people who are Journalists, social workers, health care workers, or people who are conscious of health issues. 

During the period when a noticeable number of posts were related to COVID19 (week 8 to 18), there are two small drops in the ratio of positive sentiments. One is in week 10 and another is in week 14. Both falls are followed by a significant increase in the number of COVID19 related posts. Even though these two drops are small in sentiment ratio, but the drops in the number of positive tweets were large enough to initiate triggers. It ascertains that monitoring the positive sentiment tweets can signal us the trigger in the increase of COVID19 related posts.

\vspace{-4mm}

Figure \ref{fig:geospatial_sentiment_a} shows how the ratio of the number of positive tweets vs total tweets varies in Australian states and territories. It shows that all states and territories have the positive sentiment tweet ratio of around 0.5 except Northern Territory has a slightly better ratio. This implies there is emotional stress in people over all the states and territories. However, this figure does not clearly capture the positive sentiment drop as cities are averaged over in the states and territories. In reality, some cities are affected more than others by COVID19. Therefore, we add capital cities in Figure \ref{fig:Compare_count_case_space_b} along with states and territories. 

Figure \ref{fig:geospatial_sentiment_b} shows the counts of COVID19 related tweets and positives tweets in states, territories, and capital cities. Capital cities and states that have a significant drop in positive tweet count are Sydney (syd), Melbourne (mel), Victoria (vic), and Queensland (qld). A comparison between Figure \ref{fig:geospatial_sentiment_b} and Figure \ref{fig:Compare_count_case_space_b} shows that these locations had most of the COVID19 cases. A drop in positive sentiment is correlated with the number of COVID19 cases. A drop in positive sentiment is also correlated with early mental health issues, informing that the community might need an allocation of mental health care resources in the near future. 

Two interesting facts in Figure \ref{fig:geospatial_sentiment_b} can be observed in varied behaviour between two pairs of state and its capital city, (qld, bne) and (nsw, syd) pairs. There is a significant drop in positive tweets in qld but not in bne. The majority of COVID19 cases in Queensland happened in Gold Coast and other surrounding areas instead of Brisbane. Again, there is a significant drop in positive tweet count in syd but not in nsw. The majority of COVID19 cases happened in Sydney rather than the other parts of nsw. This again emphasises that a drop in positive sentiment is directly correlated with the number of COVID19 cases.  

\vspace{-4mm}

\subsection{Topic Analysis}
\begin{figure*}[htb!]
    \centering
    
    \subfloat[Topic 0 Word Cloud]{{\includegraphics[width=2.5cm]{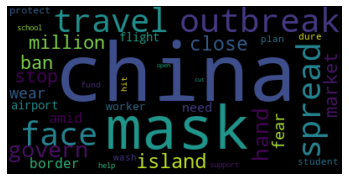}}}
    \quad
    \subfloat[Topic 0 Evolution Over Time]{{\includegraphics[width=4cm]{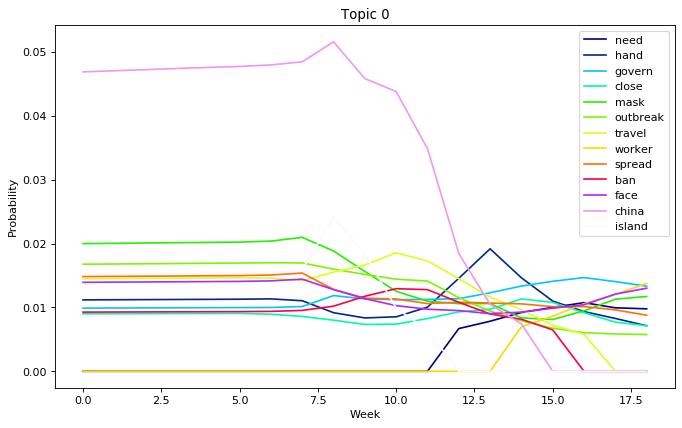}}}
    \quad
    \subfloat[Topic 1 Word Cloud]{{\includegraphics[width=2.5cm]{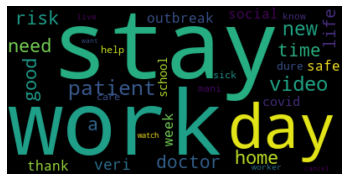}}}
    \quad
    \subfloat[Topic 1 Evolution Over Time]{{\includegraphics[width=4cm]{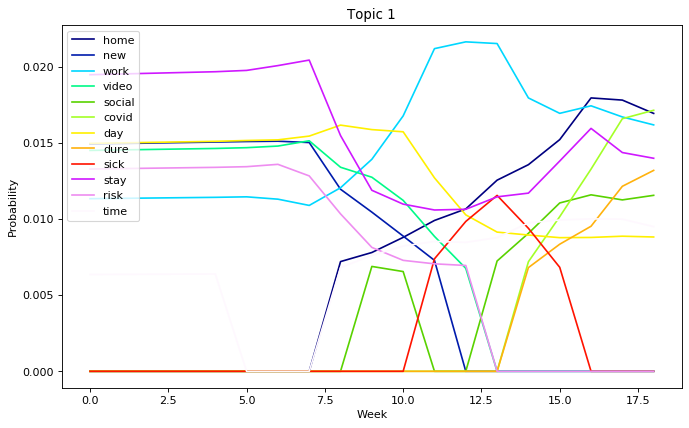}}}
    
    \subfloat[Topic 2 Word Cloud]{{\includegraphics[width=2.5cm]{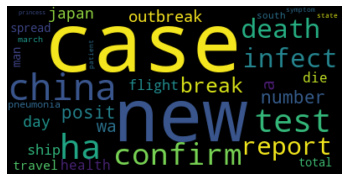}}}
    \quad
    \subfloat[Topic 2 Evolution Over Time]{{\includegraphics[width=4cm]{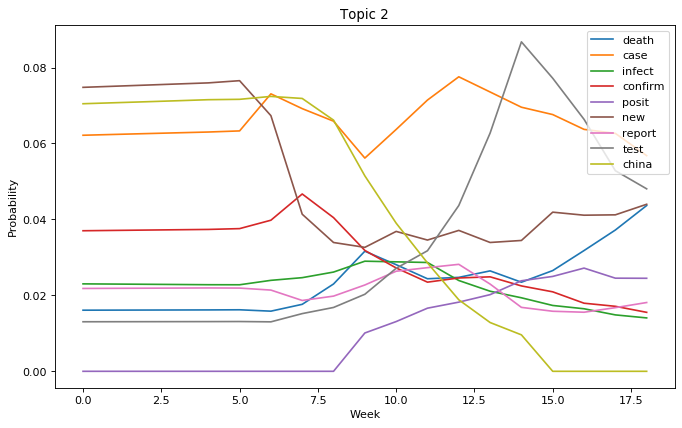}}}
    \quad
    \subfloat[Topic 3 Word Cloud]{{\includegraphics[width=2.5cm]{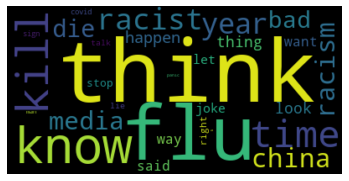}}}
    \quad
    \subfloat[Topic 3 Evolution Over Time]{{\includegraphics[width=4cm]{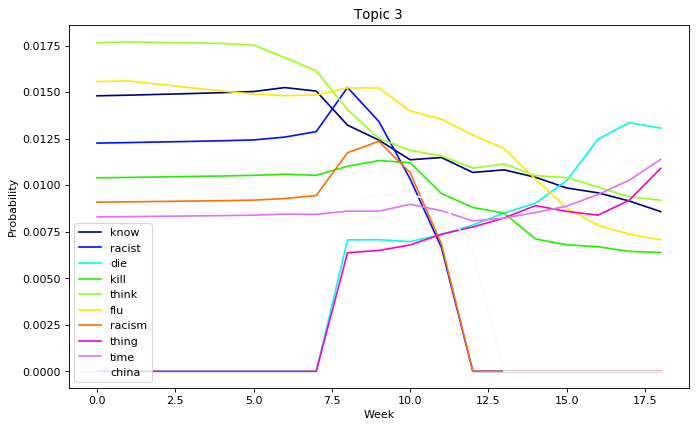}}}
    
    \subfloat[Topic 4 Word Cloud]{{\includegraphics[width=2.5cm]{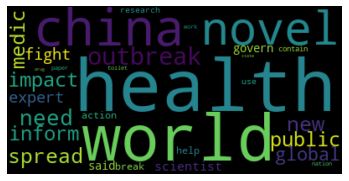}}}
    \quad
    \subfloat[Topic 4 Evolution Over Time]{{\includegraphics[width=4cm]{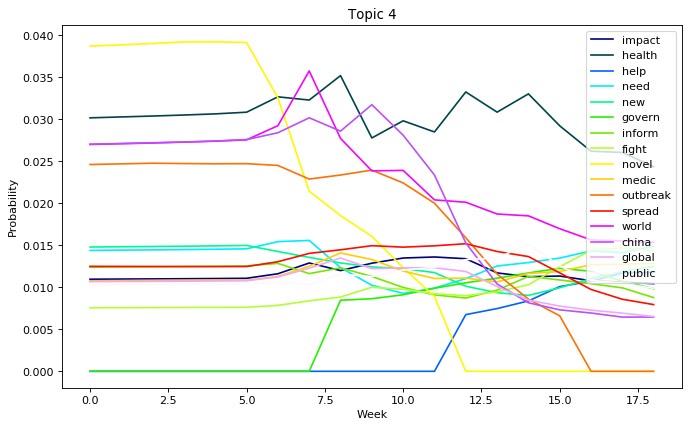}}}
    \qquad\qquad\qquad\qquad\qquad\qquad
    \subfloat[Topic Sentiment Change Over Time]{{\includegraphics[width=3.5cm]{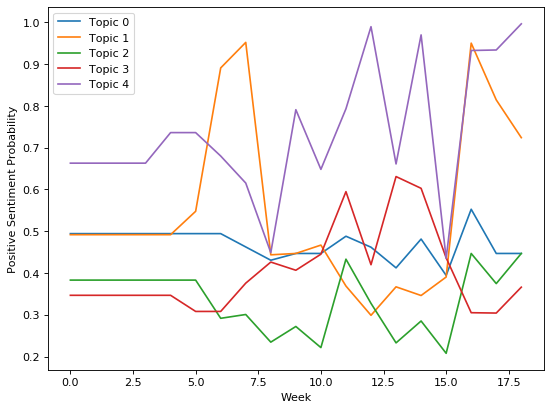}}}
    
    \caption{Topic Clouds and Topic Evolutions}%
    \label{fig:topic_evolution}%
\end{figure*}

This section shows some of the experimental results on how COVID19 related topics changed over time semantically, morphologically and sentimentally. Figure \ref{fig:topic_evolution} shows the evolution of five topics; Topic 0: controlling the spread, Topic 1: staying in isolation and working from home, Topic 2: COVID19 cases, Topic 3: racism against the Chinese community, and Topic 4: impact of COVID19 outbreak worldwide. 

Topics 0, 2 and 4 show a similar trend even though their magnitude and change rate are different. A close investigation shows that these three topics share a high similarity of subject matters. On the other hand Topics 3 and 1 do not resemble any trend. However, they somewhat inversely following each other. It is apparent that all the topics evolved over time in terms of semantics, morphology and sentiment. For example, in Topic 0 that talks about controlling the spread of coronavirus, the words `need' and `worker' newly emerged during weeks 11 and 13, whereas the words `island', `china', `travel' and `ban' lost their significance during week 12, 15, 16 and 17 respectively.  

\vspace{-3mm}
\subsection{Impact Analysis}
\begin{figure}[htb!]
    \centering
    \subfloat[Temporal Distribution of SBS]{{\includegraphics[width=5cm]{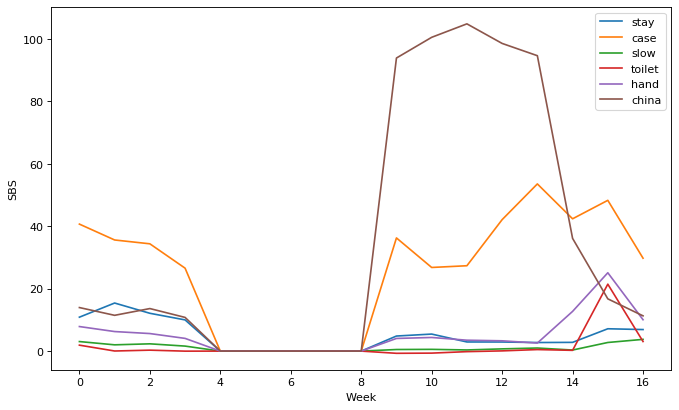}}\label{fig:sbs_short}}\hfill%
    \qquad
    \subfloat[Geospatial and Temporal Distribution of SBS for the word `China']{{\includegraphics[width=5cm]{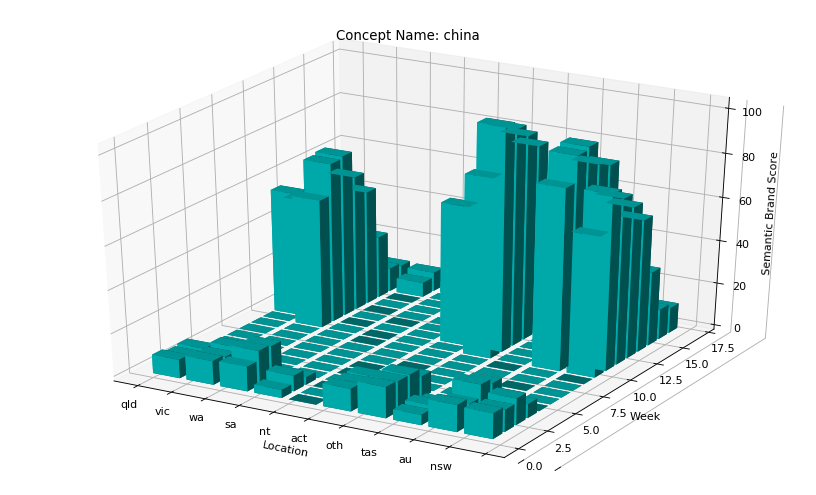}}\label{fig:sbs_china}}%
    \caption{Distribution of SBS (Impact)}%
    \label{fig:sbs}%
\end{figure}

Semantic Brand Score (SBS) can capture the impact of concepts or words in text collections that might be useful for monitoring social matters or instances. During the COVID19 pandemic, some social instances were `stay home', `positive cases', `slow the spread', `wash your hands', `toilet paper', and `China'. Figure \ref{fig:sbs_short} shows the change of SBS over time for some of the words on these instances. This figure shows that china had the highest SBS score most of the time when compared with other words. A discussion of this SBS is given below. The second highest SBS is counted for the word `case' (i.e. positive cases). This might be because people were discussing COVID19 positive cases and their implications on health, economy and jobs. Word `hand' (i.e. wash your hands) had a stable SBS score through the time except for a spike in week 15.  Word `toilet' (i.e. toilet paper) had low SBS but a spike in week 15 when there were some toilet paper related instances in Australia (e.g. toilet paper sold our in most of the stores, people fighting over buying toilet papers, etc.). 

 Figure \ref{fig:sbs_china} shows how the SBS score varied in space and time for the word China. In a certain period and some places the word China had high SBS in COVID19 related tweets. This means, China was mentioned in a lot of tweets, in a variety of topics, and a lot of topic of discussion involved the word China. One reason may be many tweets discussed the first detection of COVID19 in China. However, a high SBS also indicates that many diverse topics were influenced by the word China; and many topics (i.e. intense) were discussed in relation to the word China. This might have been influenced by the wrong assumption that China is responsible for the spreading of coronavirus as coronavirus was first detected in China. This kind of  assumption can disrupt social harmony.  As SBS can identify such incidence in space and time, it can be used for positive intervention such as providing a right information to communities and providing necessary security to the vulnerable community.

\vspace{-3mm}
\section{Discussion}
\label{sec:conclusion}
This research shows that social media data analysis is a powerful method for observing social phenomena relevant to an outbreak of an infectious disease such as COVID19. Collecting data through traditional surveys and clinical reports are time consuming and costly, and can have a time lag of few weeks between the time of medical diagnosis and the time when the data become available. Unlike traditional methods, social media data analysis is time and cost-effective that can uncover momentum and spontaneity in conversations. Besides, it can be done systematically and can be generalised a wide range of objectives.
This study analysed the discussion dynamics of COVID19 on Twitter from geospatial and temporal context using various methods. The analysis methods were found effective in capturing interesting insights and directly correlated with real-world events. 

The overall coronavirus related discussion on Twitter represent negative aspects. People were concerned about jobs, economy, and isolation in addition to health and safety. However, initiatives such as government subsidies had a positive influence on the discussion. For example, there were changes in published tweets and negative sentiments when the  new coronavirus cases were found or deaths occurred. The peaks in positive sentiment occurred during positive initiatives taken by leaders or any positive development in the health care sector. When the spread of new cases started to decline, the number of coronavirus related posts declined and the positive sentiment increased. Our observations show how social media platforms can influence the public’s risk perception, their hope and reliance on different organisational initiatives. Even it can change real-world behaviour that can have an impact on control measures enacted to mitigate an outbreak.

Dynamic topic modelling discovered a wide variety of topics in discussion, covering consequences, initiatives, impacts and peoples' behaviour with coronavirus. These topics evolved and their significance changed over time. Topic analysis provides an understanding of community discussion of COVID19 with a reasonable objectivity, precision and generality. We found that most of the COVID19 related discussion has a high concentration around a relatively small number of influential topics. For example, at the beginning most discussion was related to COVID19 outbreak in China, then the COVID19 cases in Australia and health care, then stay home and job loss. Topic modelling uncovered racism instances and SBS identified the impact in discussion. For example, our analysis showed that COVID19 pandemic created fear and that fear leaded for racism to thrive that disproportionately affecting marginalised groups.   

The findings can help government, emergency agencies, clinicians, health practitioners and caregivers to better utilise social media to understand the public opinion, sentiments, social and mental health issues related to COVID19. Such an understanding will enable proactive decision making for prioritising supports in geo-spatial locations. For example, timely disseminating and updating information related to social issues by the government can contribute to stabilising social harmony.

\vspace{-4mm}
\section{Conclusion}
Advanced analysis of social media data related a pandemic such as COVID19 is critical to protect public health, maintain social harmony and save lives. By leveraging anonymised and aggregated geo-spatial and temporal data from social media, institutions and organizations can get insights into community discussion to understand and act based on how COVID19 spread is affecting people's lives and behaviour. Specifically, the government and emergency agencies can use the insight to better understand the public opinion and sentiments to accelerate emergency responses and support post-pandemic management.

\vspace{-2mm}


\bibliographystyle{ACM-Reference-Format}
\bibliography{references.bib}

\end{document}